\documentclass[10pt]{article}
\usepackage{graphicx}
\usepackage{amsmath}
\usepackage{amssymb}
\usepackage{caption2}
\begin{document}
\bibliographystyle{prsty}
\begin{center}
{\large {\bf \sc{  Analysis of  the  $Y(4220)$ and $Y(4390)$ as   molecular  states
with  QCD sum rules }}} \\[2mm]
Zhi-Gang  Wang \footnote{E-mail: zgwang@aliyun.com.  }     \\
 Department of Physics, North China Electric Power University, Baoding 071003, P. R. China
\end{center}

\begin{abstract}
In this article,  we assign the $Y(4390)$ and  $Y(4220)$ to be the vector molecular states $D\bar{D}_1(2420)$ and $D^*\bar{D}_0^*(2400)$, respectively,    and study their masses and pole residues with the QCD sum rules in details. The present calculations only favor assigning the $Y(4390)$  to be the $D\bar{D}_1(1^{--})$ molecular state.
\end{abstract}



\section{Introduction}

In 2013, Yuan studied  the cross sections of the process  $e^+ e^- \to \pi^+\pi^- h_c$ at center-of-mass energies $3.90-4.42\,\rm{GeV}$  measured by the BESIII and the CLEO-c experiments, and observed evidence for two resonant structures,  a narrow structure of  mass  $(4216\pm 18)\,\rm{ MeV}$ and  width  $(39\pm 32) \,\rm{MeV}$, and a possible wide structure of mass $(4293\pm 9) \,\rm{MeV}$ and width $(222\pm 67) \,\rm{MeV}$ \cite{YuanCZ}.

In 2014, the BES collaboration searched for the production of $e^+e^-\to \omega\chi_{cJ}$ with $J=0,1,2,$ based on data samples collected with the BESIII detector   at   center-of-mass energies from $4.21-4.42\,\rm{GeV}$, and observed a resonance in the $\omega\chi_{c0}$ cross section,  the measured mass and width of the resonance $Y(4230)$ are $4230\pm 8\pm 6\, \rm{ MeV}$    and $ 38\pm 12\pm 2\,\rm{MeV}$, respectively \cite{BES-2014-4230}.

Recently, the BES collaboration measured the cross sections of the process $e^+ e^- \to \pi^+\pi^- h_c$   at center-of-mass energies $3.896-4.600\,\rm{GeV}$  using data samples collected with the BESIII detector, and observed two structures, the $Y(4220)$ has the mass $4218.4\pm4.0\pm0.9\,\rm{MeV}$ and with $66.0\pm9.0\pm0.4\,\rm{MeV}$ respectively, and the $Y(4390)$ has the mass $4391.6\pm6.3\pm1.0\,\rm{MeV}$ and width $139.5\pm16.1\pm0.6\,\rm{MeV}$ respectively \cite{BES-Y4390}. The $Y(4230)$ and $Y(4220)$ may be the same particle. The $Y(4230)$ has been assigned to be a vector-diquark-vector-antidiquark type vector tetraquark state \cite{Maiani-4220,Guo-4220} or a conventional meson $\psi(\rm{4S})$ \cite{Chen-4220-4S}.
The near thresholds are $M_{D^+D_1(2420)^{-}}=4293\,\rm{MeV}$, $M_{D^0D_1(2420)^{0}}=4285\,\rm{MeV}$, $M_{D^{*+}D_0^{*}(2400)^-}=4361\,\rm{MeV}$,   $M_{D^{*0}D_0^{*}(2400)^0}=4325\,\rm{MeV}$ \cite{PDG}. It is also possible to assign the $Y(4220)$ and $Y(4390)$ to be the $D\bar{D}_1(2420)$ or $D^*\bar{D}_0^*(2400)$ molecular states.

Eleven years ago, the  BaBar  collaboration observed a broad resonance ($Y(4260)$) in the initial-state radiation process
$e^+e^-\to Y(4260)\to J/\psi \pi^+\pi^-$  in the invariant-mass spectrum
of the $J/\psi \pi^+\pi^-$ \cite{BaBar-Y4260-2005}. Later, the BaBar  collaboration measured the mass and width of the $Y(4260)$ in a more precise way \cite{BaBar-Y4260-2012}.  The  cross section rises rapidly below the peak of the $Y(4260)$ and
falls more slowly above the peak \cite{PDG}. The BESIII experiment may be indicate
 that in fact the $Y(4260)$   consist of two peaks, a narrow peak around $4.22\,\rm{ GeV}$ and a wider peak around
$4.39\,\rm{GeV}$ accounting for the asymmetry.

In Ref.\cite{ZhangHuang-2009}, Zhang and Huang study the $Q\bar{q}\,\bar{Q}^{\prime}q$ type scalar, vector and axialvector molecular states with the QCD sum rules systematically by calculating the operator product expansion up to the vacuum condensates of dimension 6. The predicted molecule masses $M_{D^*\bar{D}^*_0}=4.26 \pm 0.07\,\rm{GeV}$ and $M_{D\bar{D}_1}=4.34 \pm 0.07\,\rm{GeV}$ are consistent with the $Y(4220)$ and $Y(4390)$, respectively.  However, the charge conjugations of the molecular states are not distinguished and the higher dimensional vacuum condensates are neglected.
In Ref.\cite{Lee-Nielsen},  Lee, Morita and Nielsen distinguish the  charge conjugations of the interpolating currents, calculate the operator product expansion up to the vacuum condensates of dimension 6 including the vacuum condensates of dimension 8 partly. They obtain the mass of the $D\bar{D}_1(2420)$ molecular state with $J^{PC}=1^{-+}$,  $M_{D\bar{D}_1}=4.19 \pm 0.22\,\rm{GeV}$, which differs from the prediction $M_{D\bar{D}_1}=4.34 \pm 0.07\,\rm{GeV}$ significantly.

In Refs.\cite{ZhangHuang-2009,Lee-Nielsen},  some higher dimensional  vacuum condensates involving the gluon condensate, mixed condensate and four-quark condensate are neglected. The terms associate with $\frac{1}{T^2}$, $\frac{1}{T^4}$, $\frac{1}{T^6}$ in the QCD spectral densities  manifest themselves at small values of the Borel parameter $T^2$, we have to choose large values of  $T^2$ to warrant convergence of the operator product expansion. In the Borel windows, the higher dimensional vacuum condensates  play a less important role.
The higher dimensional vacuum condensates play an important role in determining the Borel windows therefore the ground state  masses and pole residues, we should take them into account consistently.

In this article, we assign the $Y(4390)$ and  $Y(4220)$ to be  the vector molecular states $D\bar{D}_1(2420)$ and $D^*\bar{D}_0^*(2400)$, respectively,   distinguish
the charge conjugations, construct   the color singlet-singlet type   currents   to interpolate  them.
We calculate the contributions of the vacuum condensates up to dimension-10 in the operator product expansion in a consistent way, and use the energy scale formula to determine the energy scales of the QCD spectral densities \cite{WangHuang-molecule,Wang-molecule}, which differs from the routines taken in Refs.\cite{ZhangHuang-2009,Lee-Nielsen} significantly,  then  study the masses and pole residues with the QCD sum rules in details.

The article is arranged as follows:  we derive the QCD sum rules for the masses and pole residues of  the vector molecular states  in section 2; in section 3, we present the numerical results and discussions; section 4 is reserved for our conclusion.

\section{QCD sum rules for  the  vector molecular states }
In the isospin limit, the quark structures of the molecular states $D\bar{D}_1(2420)$ and $D^*\bar{D}_0^*(2400)$ can be symbolically written as
\begin{eqnarray}
\bar{u}d \bar{c}c\, ,\,\,\, \frac{\bar{u}u-\bar{d}d }{\sqrt{2}}\bar{c}c\, ,\,\,\,\bar{d}u \bar{c}c\, ,\,\,\,\frac{\bar{u}u+\bar{d}d }{\sqrt{2}}\bar{c}c\, .
\end{eqnarray}
The isospin triplet $\bar{u}d \bar{c}c$, $\frac{\bar{u}u-\bar{d}d }{\sqrt{2}}\bar{c}c$, $\bar{d}u \bar{c}c$ and isospin singlet $\frac{\bar{u}u+\bar{d}d }{\sqrt{2}}\bar{c}c$ have degenerate masses. In this article, we  take the isospin limit and study the masses of the charged partners of the $Y(4220)$ and $Y(4390)$ for simplicity.

In the following, we write down  the two-point correlation functions $\Pi_{\mu\nu}(p)$  in the QCD sum rules,
\begin{eqnarray}
\Pi_{\mu\nu}(p)&=&i\int d^4x e^{ip \cdot x} \langle0|T\left\{J_\mu(x)J_\nu^{\dagger}(0)\right\}|0\rangle \, ,
\end{eqnarray}
where $J_\mu(x)=J_\mu^1(x),\,J_\mu^2(x),\,J_\mu^3(x),\,J_\mu^4(x)$,
\begin{eqnarray}
J^1_\mu(x)&=&\frac{1}{\sqrt{2}}\left\{ \bar{u}(x)i\gamma_5c(x)\bar{c}(x)\gamma_\mu \gamma_5 d(x)-\bar{u}(x)\gamma_\mu \gamma_5c(x)\bar{c}(x)i\gamma_5 d(x)\right\} \, , \nonumber \\
J^2_\mu(x)&=&\frac{1}{\sqrt{2}}\left\{ \bar{u}(x)i\gamma_5c(x)\bar{c}(x)\gamma_\mu \gamma_5 d(x)+\bar{u}(x)\gamma_\mu \gamma_5c(x)\bar{c}(x)i\gamma_5 d(x)\right\} \, , \nonumber \\
J^3_\mu(x)&=&\frac{1}{\sqrt{2}}\left\{ \bar{u}(x)c(x)\bar{c}(x)\gamma_\mu  d(x)+\bar{u}(x)\gamma_\mu c(x)\bar{c}(x) d(x)\right\} \, , \nonumber \\
J^4_\mu(x)&=&\frac{1}{\sqrt{2}}\left\{ \bar{u}(x)c(x)\bar{c}(x)\gamma_\mu  d(x)-\bar{u}(x)\gamma_\mu c(x)\bar{c}(x) d(x)\right\} \, ,
\end{eqnarray}
 Under charge conjugation transform $\widehat{C}$, the currents $J_\mu(x)$ have the properties,
\begin{eqnarray}
\widehat{C}J^{1/3}_{\mu}(x)\widehat{C}^{-1}&=& - J^{1/3}_\mu(x) |_{u\leftrightarrow d} \, , \nonumber\\
\widehat{C}J^{2/4}_{\mu}(x)\widehat{C}^{-1}&=&+ J^{2/4}_\mu(x)|_{u\leftrightarrow d} \, .
\end{eqnarray}

The charge conjugations of the molecular states $Y(4220)$ and $Y(4390)$ are unknown. If the decays take place through
 \begin{eqnarray}
 Y(4220/4390)&\to& \rho h_c \to h_c \pi^+ \pi^- \, ,
 \end{eqnarray}
 the charge conjugation is positive; on the other hand, if the decays take place
 through
 \begin{eqnarray}
 Y(4220/4390)&\to& Z_c^{\pm}(4025) \pi^{\mp} \to h_c \pi^+ \pi^- \, ,
 \end{eqnarray}
 the charge conjugation is negative, where we assume that there is a relative S-wave between the intermediate mesons  $\rho h_c$ or $Z_c^{\pm}(4025) \pi^{\mp}$.
 The decay
 \begin{eqnarray}
 Y(4230) &\to& \omega \chi_{c0} \, ,
 \end{eqnarray}
 has been observed \cite{BES-2014-4230}. If the $Y(4220)$ and $Y(4230)$ are the same particle, the $Y(4220)$ maybe have the quantum numbers $J^{PC}=1^{--}$.

At the phenomenological side, we insert  a complete set of intermediate hadronic states with
the same quantum numbers as the current operators $J_\mu(x)$ into the
correlation functions $\Pi_{\mu\nu}(p)$  to obtain the hadronic representation
\cite{SVZ79,Reinders85}. After isolating the ground state
contributions of the vector molecular  states, we get the following results,
\begin{eqnarray}
\Pi_{\mu\nu}(p)&=&\frac{\lambda_{Y}^2}{M_{Y}^2-p^2}\left(-g_{\mu\nu} +\frac{p_\mu p_\nu}{p^2}\right) +\cdots \, \, ,
\end{eqnarray}
where the pole residues  $\lambda_{Y}$ are defined by $ \langle 0|J_\mu(0)|Y(p)\rangle=\lambda_{Y} \,\varepsilon_\mu$,
the $\varepsilon_\mu$ are the polarization vectors of the  vector molecular states.

In the following, we  perform Fierz re-arrangement for the currents $J_\mu$  both in the color space and Dirac-spinor  space  to obtain the  results,
\begin{eqnarray}
J_\mu^1&=&\frac{1}{2\sqrt{2}}\left\{\frac{1}{3} i\bar{u}\gamma_\mu d\, \bar{c}c -\frac{1}{3}i\bar{u} d\, \bar{c}\gamma_\mu c -\frac{1}{3}\bar{u}\gamma^\beta\gamma_5 d\, \bar{c}\sigma_{\mu\beta}\gamma_5 c +\frac{1}{3}\bar{u}\sigma_{\mu\beta} \gamma_5d\, \bar{c}\gamma^\beta\gamma_5 c\right.   \nonumber\\
&&\left.+ \frac{1}{2}i\bar{u}\gamma_\mu \lambda^ad\, \bar{c}\lambda^ac -\frac{1}{2}i\bar{u} \lambda^ad\, \bar{c}\gamma_\mu \lambda^ac -\frac{1}{2}\bar{u}\gamma^\beta\gamma_5 \lambda^ad\, \bar{c}\sigma_{\mu\beta}\gamma_5 \lambda^ac +\frac{1}{2}\bar{u}\sigma_{\mu\beta} \gamma_5\lambda^ad\, \bar{c}\gamma^\beta\gamma_5 \lambda^ac\right\}\, , \nonumber\\
J_\mu^2&=&\frac{1}{2\sqrt{2}}\left\{ \frac{1}{3}\bar{u}\sigma_{\mu\beta} d\, \bar{c}\gamma^\beta c+\frac{1}{3}\bar{u}\gamma^\beta d\, \bar{c}\sigma_{\mu\beta} c-\frac{1}{3}\bar{u}i\gamma_5 d\, \bar{c}\gamma_\mu\gamma_5c -\frac{1}{3}\bar{u}\gamma_\mu\gamma_5 d\, \bar{c}i\gamma_5 c\right. \nonumber\\
&&+\left. \frac{1}{2}\bar{u}\sigma_{\mu\beta}\lambda^a d\, \bar{c}\gamma^\beta\lambda^a c+\frac{1}{2}\bar{u}\gamma^\beta \lambda^ad\, \bar{c}\sigma_{\mu\beta} \lambda^ac-\frac{1}{2}\bar{u}i\gamma_5 \lambda^ad\, \bar{c}\gamma_\mu\gamma_5\lambda^ac -\frac{1}{2}\bar{u}\gamma_\mu\gamma_5 \lambda^ad\, \bar{c}i\gamma_5 \lambda^ac\right\} \, , \nonumber\\
J_\mu^3&=&\frac{1}{2\sqrt{2}}\left\{ -\frac{1}{3}\bar{u}\gamma_\mu d\, \bar{c}c -\frac{1}{3}\bar{u} d\, \bar{c}\gamma_\mu c -\frac{1}{3}i\bar{u}\gamma^\beta\gamma_5 d\, \bar{c}\sigma_{\mu\beta}\gamma_5 c -\frac{1}{3}i\bar{u}\sigma_{\mu\beta} \gamma_5d\, \bar{c}\gamma^\beta\gamma_5 c\right. \nonumber\\
&&\left. -\frac{1}{2}\bar{u}\gamma_\mu \lambda^ad\, \bar{c}\lambda^ac -\frac{1}{2}\bar{u} \lambda^ad\, \bar{c}\gamma_\mu \lambda^ac -\frac{1}{2}i\bar{u}\gamma^\beta\gamma_5 \lambda^ad\, \bar{c}\sigma_{\mu\beta}\gamma_5 \lambda^ac -\frac{1}{2}i\bar{u}\sigma_{\mu\beta} \gamma_5\lambda^ad\, \bar{c}\gamma^\beta\gamma_5 \lambda^ac\right\} \, , \nonumber\\
J_\mu^4&=&\frac{1}{2\sqrt{2}}\left\{ -\frac{1}{3}i\bar{u}\sigma_{\mu\beta} d\, \bar{c}\gamma^\beta c+\frac{1}{3}i\bar{u}\gamma^\beta d\, \bar{c}\sigma_{\mu\beta} c+\frac{1}{3}i\bar{u}i\gamma_5 d\, \bar{c}\gamma_\mu\gamma_5c -\frac{1}{3}i\bar{u}\gamma_\mu\gamma_5 d\, \bar{c}i\gamma_5 c\right. \nonumber\\
&&\left. -\frac{1}{2}i\bar{u}\sigma_{\mu\beta}\lambda^a d\, \bar{c}\gamma^\beta \lambda^ac+\frac{1}{2}i\bar{u}\gamma^\beta \lambda^ad\, \bar{c}\sigma_{\mu\beta} \lambda^ac+\frac{1}{2}i\bar{u}i\gamma_5 \lambda^ad\, \bar{c}\gamma_\mu\gamma_5\lambda^ac -\frac{1}{2}i\bar{u}\gamma_\mu\gamma_5 \lambda^ad\, \bar{c}i\gamma_5 \lambda^ac\right\} \, . \nonumber\\
\end{eqnarray}
The components $\bar{u}\Gamma d\, \bar{c}\Gamma^\prime c$ and
$\bar{u}\Gamma \lambda^ad\, \bar{c}\Gamma^\prime \lambda^ac$ couple potentially to a series of charmonium-light-meson pairs or charmonium-like molecular states or charmonium-like molecule-like states, where $\Gamma, \Gamma^\prime=1,\,\gamma_\mu,\, \gamma_\mu\gamma_5,\, i\gamma_5,\,\sigma_{\mu\beta}, \, \sigma_{\mu\beta}\gamma_5 $. For example, the current $J^1_\mu$ couples potentially to the meson pairs through its components,
\begin{eqnarray}
  \bar{u}\gamma_\mu d\, \bar{c}c &\propto&  \chi_{c0}\rho^- \, ,\,\cdots\, , \nonumber\\
  \bar{u} d\, \bar{c}\gamma_\mu c &\propto&  J/\psi a_0^{-}(980) \, ,\, \cdots\, , \nonumber\\
  \bar{u}\gamma^\beta\gamma_5 d\, \bar{c}\sigma_{\mu\beta}\gamma_5 c &\propto& J/\psi a_1^-(1260),\,\, J/\psi \pi^-, \,\,h_c a_1^-(1260),\,\, h_c \pi^- \, ,\,\cdots\,, \nonumber\\
  \bar{u}\sigma_{\mu\beta} \gamma_5d\, \bar{c}\gamma^\beta\gamma_5 c &\propto& \eta_c\rho^-, \,\, \chi_{c1}\rho^-, \,\, \eta_c h_1^-(1170), \,\, \chi_{c1}h_1^-(1170)\, ,\,\cdots\, .
\end{eqnarray}
We cannot distinguish those
contributions  to study them exclusively, and   assume that the  currents  $\bar{u}\Gamma d\, \bar{c}\Gamma^\prime c$ and
$\bar{u}\Gamma \lambda^ad\, \bar{c}\Gamma^\prime \lambda^ac$
 couples to a particular resonance $Y$,   which is a special  superposition of the scattering states,
molecular states and molecule-like states, and embodies   the  net
effects. Some meson pairs (in other words, its components) such as  $\chi_{c0}\rho$, $J/\psi a_0(980)$, $J/\psi \pi$, $\cdots$ lie below the $Y$, the $Y$ can decay to those meson pairs easily through fall-apart mechanism,  although the rearrangements in the   color space and Dirac-spinor  space are highly non-trivial, the decays contribute   a finite width to the $Y$.

 In the following, we
study the contributions of the  intermediate   meson-loops to the correlation function $\Pi_{\mu\nu}(p)$ for the current $J_\mu^1(x)$ as an example, the   current $J_\mu^1(x)$  has nonvanishing couplings  with the scattering states   $J/\psi a_0(980)$, $\chi_{c0} \rho$, etc.
\begin{eqnarray}
\Pi_{\mu\nu}(p)&=&-\frac{\widehat{\lambda}_{Y}^{2}}{ p^2-\widehat{M}_{Y}^2-\Sigma_{J/\psi a_0(980)}(p)-\Sigma_{\chi_{c0} \rho}(p)+\cdots}\left(g_{\mu\nu} -\frac{p_\mu p_\nu}{p^2}\right)+\cdots \, ,
\end{eqnarray}
where the $\widehat{\lambda}_{Y}$ and $\widehat{M}_{Y}$ are bare quantities to absorb the divergences in the self-energies $\Sigma_{J/\psi a_0(980)}(p)$, $\Sigma_{\chi_{c0} \rho}(p)$, etc.
The renormalized self-energies  contribute  a finite imaginary part to modify the dispersion relation,
\begin{eqnarray}
\Pi_{\mu\nu}(p) &=&-\frac{\lambda_{Y}^{2}}{ p^2-M_{Y}^2+i\sqrt{p^2}\Gamma(p^2)}\left(g_{\mu\nu} -\frac{p_\mu p_\nu}{p^2}\right)+\cdots \, .
 \end{eqnarray}
The physical  widths $\Gamma_{Y(4220)}=66.0\pm9.0\pm0.4\,\rm{MeV}$ and $M_{Y(4390)}=139.5\pm16.1\pm0.6\,\rm{MeV}$ \cite{BES-Y4390} are not large, the finite width effects can be absorbed into the pole residues $\lambda_{Y}$. In previous works, we observed that the effects of the  finite widths, such as $\Gamma_{X(4500)}=92 \pm 21 ^{+21}_{-20} \,\rm{MeV}$, $\Gamma_{X(4700)}= 120 \pm 31 ^{+42}_{-33} \,\rm{MeV}$, $\Gamma_{Z_c(4200)} = 370^{+70}_{-70}{}^{+70}_{-132}\,\rm{MeV}$, can be safely absorbed into the pole residues $\lambda_{X/Z_c}$ \cite{Wang-Zc4200}. In this article, we take the zero width approximation, and expect that the predicted masses are reasonable.

 We carry out the operator product expansion  in a consistent way, and obtain the QCD spectral densities  through dispersion relation, then
 we take the
quark-hadron duality below the continuum thresholds $s_0$ and perform Borel transform  with respect to
the variable $P^2=-p^2$ to obtain  the following QCD sum rules,
\begin{eqnarray}
\lambda^2_{Y}\, \exp\left(-\frac{M^2_{Y}}{T^2}\right)= \int_{4m_c^2}^{s_0} ds\, \rho(s) \, \exp\left(-\frac{s}{T^2}\right) \, ,
\end{eqnarray}
where $\rho(s)=\rho_1(s),\,\rho_2(s),\,\rho_3(s),\,\rho_4(s)$,
\begin{eqnarray}
\rho_1(s)&=&\rho(s,r)\mid_{r=1}\, , \nonumber\\
\rho_2(s)&=&\rho(s,r)\mid_{r=-1}\, , \nonumber\\
\rho_3(s)&=&\rho(s,r)\mid_{r=1,m_c \to -m_c}\, , \nonumber\\
\rho_4(s)&=&\rho(s,r)\mid_{r=-1,m_c \to -m_c}\, ,
\end{eqnarray}
the explicit expressions of the QCD spectral densities $\rho(s,r)$ are given in the Appendix.
 In this article, we carry out the
operator product expansion to the vacuum condensates  up to dimension-10 and assume vacuum saturation for the  higher dimension vacuum condensates.
 The condensates $\langle g_s^3 GGG\rangle$, $\langle \frac{\alpha_s GG}{\pi}\rangle^2$,
 $\langle \frac{\alpha_s GG}{\pi}\rangle\langle \bar{q} g_s \sigma Gq\rangle$ have the dimensions 6, 8, 9 respectively,  but they are   the vacuum expectations
of the operators of the order    $\mathcal{O}( \alpha_s^{3/2})$, $\mathcal{O}(\alpha_s^2)$, $\mathcal{O}( \alpha_s^{3/2})$ respectively, and discarded.  We take
the truncations $n\leq 10$ and $k\leq 1$ in a consistent way,
the operators of the orders $\mathcal{O}( \alpha_s^{k})$ with $k> 1$ are  discarded \cite{WangHuang-molecule,Wang-molecule,WangHuang-Tetraquark}. Furthermore,  the numerical values of the  condensates $\langle g_s^3 GGG\rangle$, $\langle \frac{\alpha_s GG}{\pi}\rangle^2$,
 $\langle \frac{\alpha_s GG}{\pi}\rangle\langle \bar{q} g_s \sigma Gq\rangle$   are very small, and  they are neglected safely.

 We derive    Eq.(13) with respect to  $\tau=\frac{1}{T^2}$, and eliminate the
 pole residues  $\lambda_{Y}$ to obtain the QCD sum rules for the masses,
 \begin{eqnarray}
 M^2_{Y}= \frac{\int_{4m_c^2}^{s_0} ds\left(-\frac{d}{d \tau }\right)\rho(s)e^{-\tau s}}{\int_{4m_c^2}^{s_0} ds \rho(s)e^{-\tau s}}\, .
\end{eqnarray}

\section{Numerical results and discussions}
The vacuum condensates are taken to be the standard values
$\langle\bar{q}q \rangle=-(0.24\pm 0.01\, \rm{GeV})^3$,
$\langle\bar{q}g_s\sigma G q \rangle=m_0^2\langle \bar{q}q \rangle$,
$m_0^2=(0.8 \pm 0.1)\,\rm{GeV}^2$, $\langle \frac{\alpha_s
GG}{\pi}\rangle=(0.33\,\rm{GeV})^4 $    at the energy scale  $\mu=1\, \rm{GeV}$
\cite{SVZ79,Reinders85,ColangeloRV}.
The quark condensate and mixed quark condensate evolve with the   renormalization group equation,
$\langle\bar{q}q \rangle(\mu)=\langle\bar{q}q \rangle(Q)\left[\frac{\alpha_{s}(Q)}{\alpha_{s}(\mu)}\right]^{\frac{4}{9}}$,
 and $\langle\bar{q}g_s \sigma Gq \rangle(\mu)=\langle\bar{q}g_s \sigma Gq \rangle(Q)\left[\frac{\alpha_{s}(Q)}{\alpha_{s}(\mu)}\right]^{\frac{2}{27}}$.
In the article, we take the $\overline{MS}$ mass $m_{c}(m_c)=(1.275\pm0.025)\,\rm{GeV}$  from the Particle Data Group \cite{PDG} and take into account
the energy-scale dependence of  the $\overline{MS}$ mass,
\begin{eqnarray}
m_c(\mu)&=&m_c(m_c)\left[\frac{\alpha_{s}(\mu)}{\alpha_{s}(m_c)}\right]^{\frac{12}{25}} \, ,\nonumber\\
\alpha_s(\mu)&=&\frac{1}{b_0t}\left[1-\frac{b_1}{b_0^2}\frac{\log t}{t} +\frac{b_1^2(\log^2{t}-\log{t}-1)+b_0b_2}{b_0^4t^2}\right]\, ,
\end{eqnarray}
  where $t=\log \frac{\mu^2}{\Lambda^2}$, $b_0=\frac{33-2n_f}{12\pi}$, $b_1=\frac{153-19n_f}{24\pi^2}$, $b_2=\frac{2857-\frac{5033}{9}n_f+\frac{325}{27}n_f^2}{128\pi^3}$,  $\Lambda=213\,\rm{MeV}$, $296\,\rm{MeV}$  and  $339\,\rm{MeV}$ for the flavors  $n_f=5$, $4$ and $3$, respectively  \cite{PDG}.

The hidden charm (or hidden bottom) four-quark systems  $Qq\bar{Q}\bar{q}^\prime$ could be described
by a double-well potential in the heavy quark limit. The  heavy quark $Q$ serves as one  static well potential and  combines with the light antiquark $\bar{q}^\prime$  to form a heavy meson-like state or correlation (not a physical meson) in  color singlet. The  heavy antiquark $\bar{Q}$ serves  as the other  static well potential and combines with the light quark state $q$  to form another heavy meson-like state or correlation (not a physical meson) in  color singlet.    The two meson-like states (not two physical mesons)  combine together to form a  physical molecular state.  Then the double  heavy molecular state  $Y$ is characterized by the effective heavy quark mass ${\mathbb{M}}_Q$ and the virtuality $V=\sqrt{M^2_{Y}-(2{\mathbb{M}}_Q)^2}$ \cite{WangHuang-molecule,Wang-molecule}. It is natural to choose the energy scales of the QCD spectral densities as $\mu=V$, which works well in the QCD sum rules for the molecular states. In Ref.\cite{WangHuang-molecule}, we obtain the optimal value  ${\mathbb{M}}_c=1.84\,\rm{GeV}$. Recently, we re-checked the numerical calculations and corrected  a small error involving the mixed condensates. After the small error was corrected,
the Borel windows are modified slightly and the predictions are also improved slightly, but the conclusions survive. In this article, we choose the updated value ${\mathbb{M}}_c=1.85\,\rm{GeV}$.

In the  scenario of  molecular states, we study the color singlet-singlet type and  octet-octet  type scalar, axial-vector and tensor  hadronic molecular states with the QCD sum rules in a systematic way \cite{WangHuang-molecule,Wang-molecule}, and assign
the $X(3872)$, $Z_c(3900/3885)$,   $Y(4140)$, $Z_c(4020/4025)$ and $Z_b(10610/10650)$ to be the molecular states tentatively,
\begin{eqnarray}
X(3872)&=&\frac{1}{\sqrt{2}}\left( D\overline{D}^{*} -  D^{*}\overline{D}\right) \,\,\,({\rm with}\,\,\,1^{++})\, , \nonumber \\
Z_c(3900/3885)&=&\frac{1}{\sqrt{2}}\left( D\overline{D}^{*} +  D^{*}\overline{D}\right)\,\,\,({\rm with}\,\,\,1^{+-}) \, , \nonumber\\
Z_c(4020/4025)&=& D^*\overline{D}^{*} \,\,\,({\rm with}\,\,\,1^{+-}\,\,\,{\rm or}\,\,\,2^{++})  \, ,\nonumber \\
Y(4140)&=&D_s^*\overline{D}_s^* \,\,\,({\rm with}\,\,\, 0^{++})\, , \nonumber\\
Z_b(10610)&=&\frac{1}{\sqrt{2}}\left( B\overline{B}^* + B^*\overline{B}\right)\,\,\,({\rm with}\,\,\,1^{+-}) \, ,\nonumber \\
Z_b(10650)&=&  B^{*}\overline{B}^{*} \,\,\,({\rm with}\,\,\,1^{+-}) \, .
\end{eqnarray}

 Now we search for  the  Borel parameters $T^2$ and continuum threshold
parameters $s_0$  to satisfy the  following four criteria:

$\bullet$ Pole dominance at the phenomenological side;

$\bullet$ Convergence of the operator product expansion;

$\bullet$ Appearance of the Borel platforms;

$\bullet$ Satisfying the energy scale formula.

The resulting Borel parameters, continuum threshold parameters, pole contributions and energy scales are shown explicitly in Table 1. From the Table, we can
see that the central values of the pole contributions are larger than $50\%$,
the pole dominance condition can be  satisfied.
In the Borel windows, the  contributions come from the vacuum condensates $D_i$ of dimension $i$  are
\begin{eqnarray}
D\bar{D}_1 (1^{--})&:& D_{0}=(94-95)\%\, ,\,\, D_3=0\%\, ,\,\, D_4\ll 1\%\, ,\,\,D_5=(20-24)\% \, ,\nonumber\\
&&  D_6=-(12-17)\%\, ,\,\, D_7=-(1-2)\%\, ,\,\, -D_8< 1\%\, ,\,\, D_{10}\ll 1\% \, , \nonumber\\
D\bar{D}_1 (1^{-+})&:& D_{0}=(122-127)\%\, ,\,\, D_3=-(24-27)\%\, ,\,\, D_4=- 1\%\, ,\,\,D_5=(24-29)\%\, ,\nonumber\\
&& D_6=-(19-27)\%\, ,\,\, D_7=-(2-3)\%\, ,\,\, D_8\leq 1\%\, ,\,\, D_{10}\ll 1\% \, , \nonumber\\
D^*\bar{D}_0^* (1^{--})&:& D_{0}=(118-122)\%\, ,\,\, D_3=0\%\, ,\,\, D_4\ll 1\%\, ,\,\,D_5=-(13-16)\%\, , \nonumber\\
&& D_6=-(5-7)\%\, , \,\,D_7<1\%\, ,\,\, -D_8< 1\%\, ,\,\, D_{10}\ll 1\% \, , \nonumber\\
D^*\bar{D}_0^* (1^{-+})&:& D_{0}=(111-116)\%\, ,\,\, D_3=(19-22)\%\, ,\,\, D_4=- 1\%\, ,\,\,D_5=-(18-22)\%\, ,\nonumber\\
&& D_6=-(13-18)\%\, ,\,\, D_7=(1-2)\%\, ,\,\, D_8< 1\%\, ,\,\, D_{10}\ll 1\% \, ,
\end{eqnarray}
where $i=0$, $3$, $4$, $5$, $6$, $7$, $8$, $10$. The operator product expansion is well convergent. In the QCD sum rules for the hidden charm  tetraquark states and molecular states, the operator product expansion converges slowly,  we have to postpone the Borel parameters to large values. Larger Borel parameters lead to smaller pole contributions at the hadron side. So in the QCD sum rules for the hidden charm tetraquark states and molecular states, the Borel windows are rather small, $T_{max}^2-T^2_{min}\approx 0.4\,\rm{GeV}^2$, while the low bounds  of the  pole contributions are about $(40-45)\%$. From Table 1, we can see that the threshold parameters and the predicted masses  satisfy the relation $\sqrt{s_0}=M_{Y}+(0.4\sim 0.6)\,\rm{GeV}$. Naively, we expect that the energy gap between the ground state and the first radial excited state is about $0.4\sim0.6\,\rm{GeV}$, the present predictions are reasonable. Although the low bounds  of the  pole contributions   are less than $50\%$, the contaminations of the radial excited states and continuum states are expected to be excluded by the continuum threshold parameters $s_0$.

\begin{table}
\begin{center}
\begin{tabular}{|c|c|c|c|c|c|c|c|}\hline\hline
                            &$T^2(\rm{GeV}^2)$ &$\sqrt{s_0}(\rm{GeV})$ &pole        &$\mu(\rm{GeV})$  &$M_{Y}(\rm{GeV})$ &$\lambda_{Y}(10^{-2}\rm{GeV}^5)$\\ \hline
$D\bar{D}_1$ ($1^{--}$)     &$3.2-3.6$         &$4.9\pm0.1$            &$(45-65)\%$ &$2.3$            &$4.36\pm0.08$     &$3.97\pm 0.54 $    \\ \hline
$D\bar{D}_1$ ($1^{-+}$)     &$3.5-3.9$         &$5.1\pm0.1$            &$(44-63)\%$ &$2.7$            &$4.60\pm 0.08$    &$5.26 \pm0.65$     \\ \hline
$D^*\bar{D}_0^*$ ($1^{--}$) &$4.0-4.4$         &$5.3\pm0.1$            &$(44-61)\%$ &$3.0$            &$4.78\pm0.07$     &$7.56\pm 0.84$    \\ \hline
$D^*\bar{D}_0^*$ ($1^{-+}$) &$3.8-4.2$         &$5.2\pm0.1$            &$(44-61)\%$ &$2.9$            &$4.73\pm0.07$     &$6.83\pm 0.84$    \\ \hline \hline
\end{tabular}
\end{center}
\caption{ The Borel parameters, continuum threshold parameters, pole contributions, energy scales, masses and pole residues of the vector molecular  states. }
\end{table}

\begin{figure}
\centering
\includegraphics[totalheight=6cm,width=7cm]{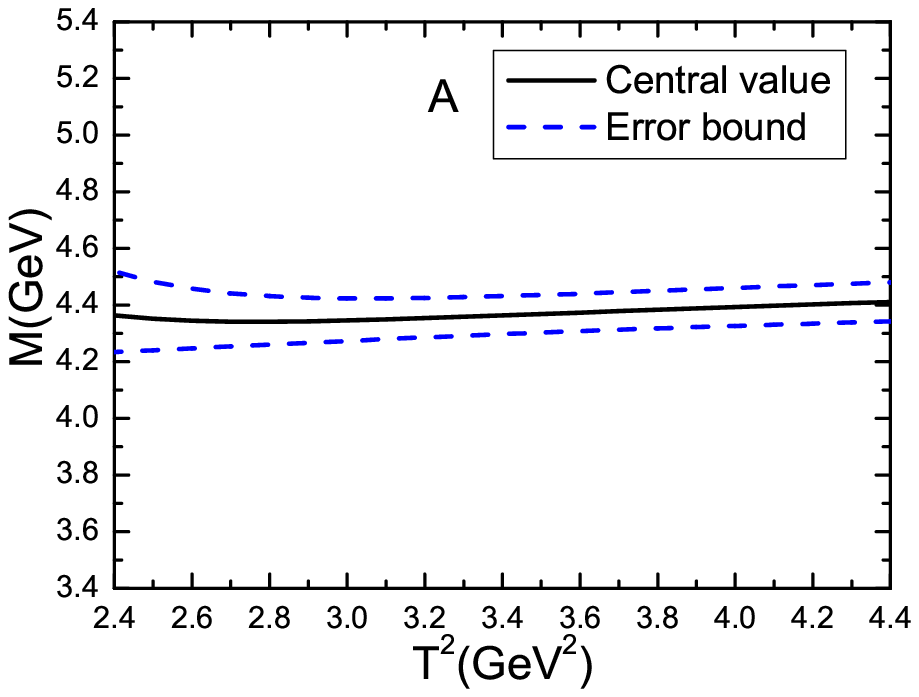}
\includegraphics[totalheight=6cm,width=7cm]{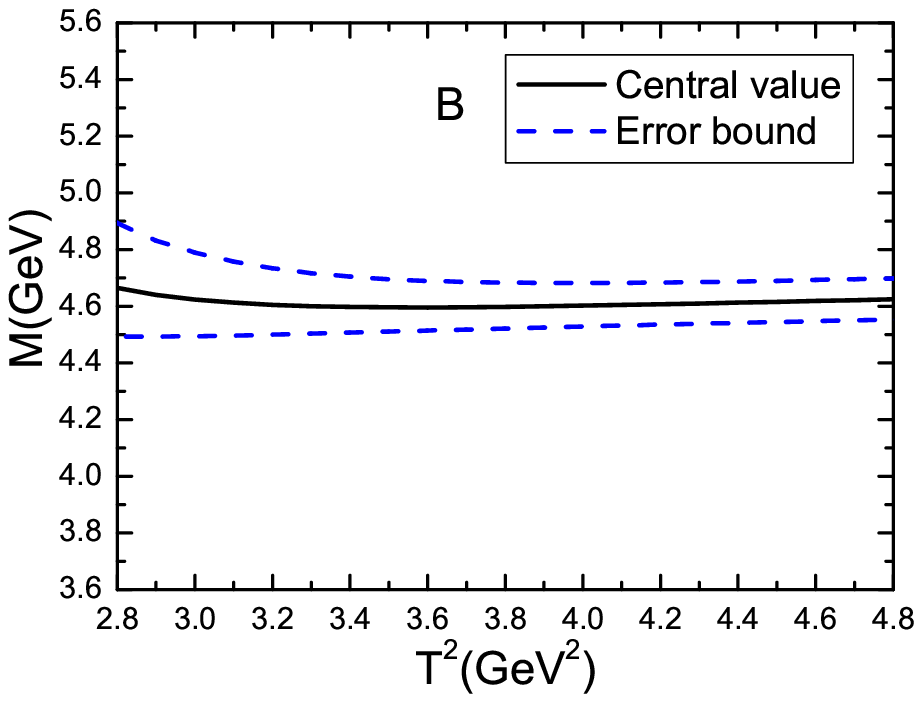}
\includegraphics[totalheight=6cm,width=7cm]{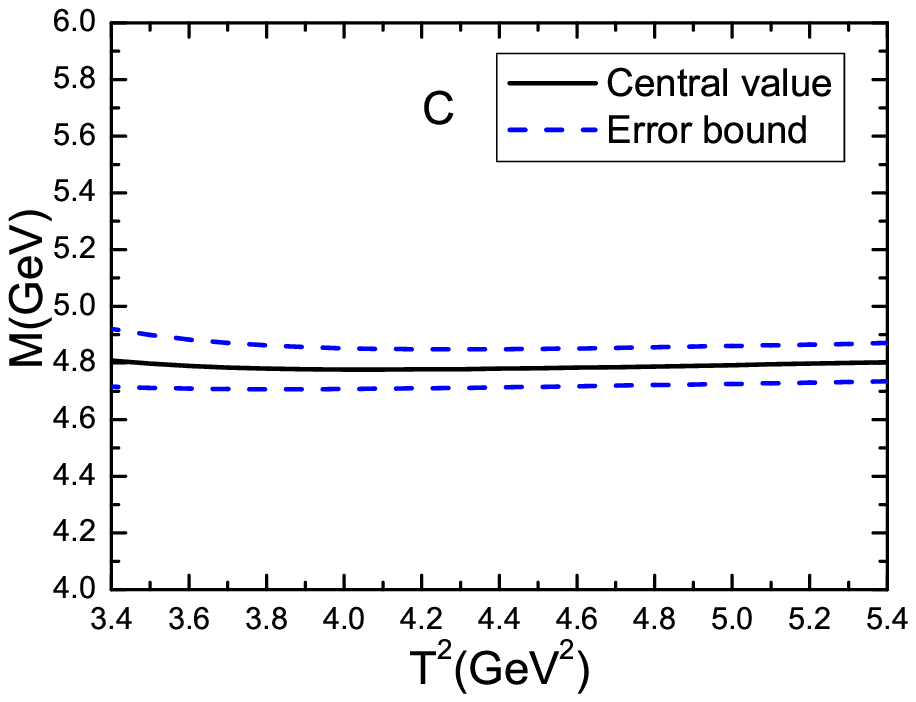}
\includegraphics[totalheight=6cm,width=7cm]{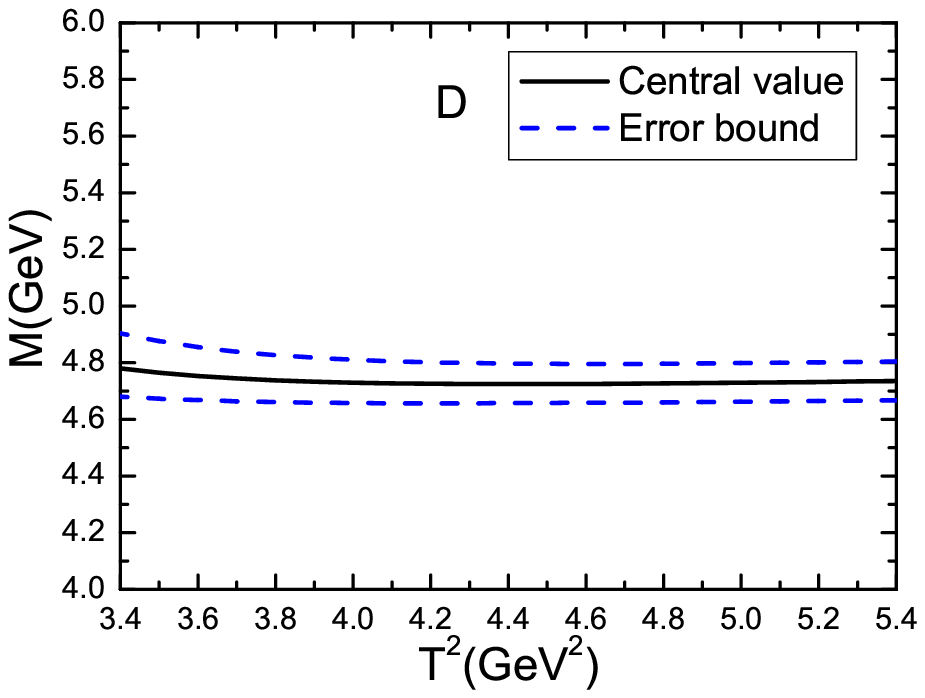}
  \caption{ The masses  with variations of the  Borel parameters $T^2$, where  the $A$, $B$, $C$ and $D$ denote the molecular states
  $D\bar{D}_1(1^{--})$, $D\bar{D}_1(1^{-+})$, $D^*\bar{D}_0^*(1^{--})$ and $D^*\bar{D}_0^*(1^{-+})$, respectively. }
\end{figure}

\begin{figure}
\centering
\includegraphics[totalheight=6cm,width=7cm]{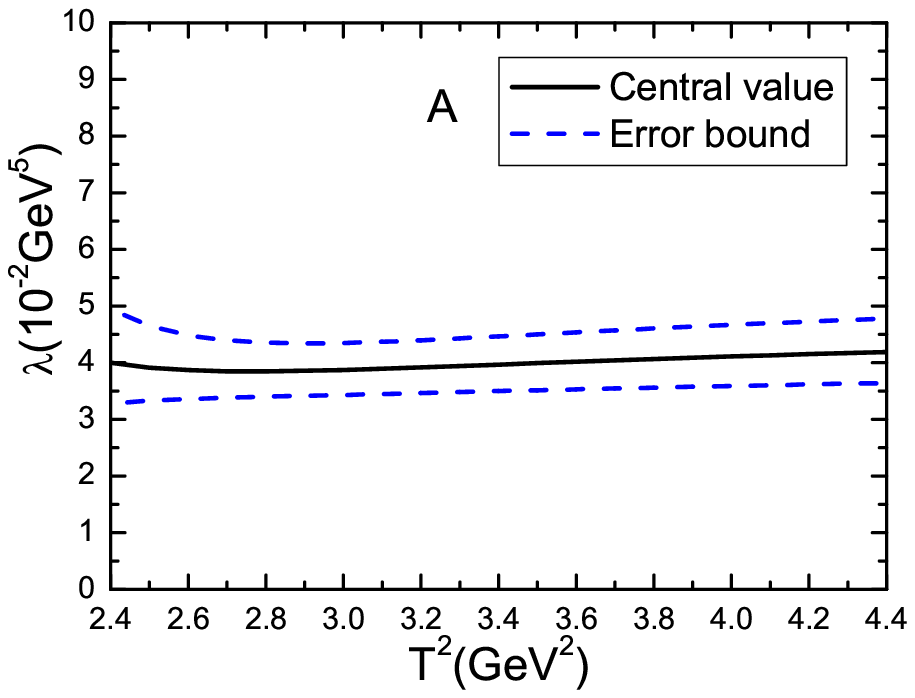}
\includegraphics[totalheight=6cm,width=7cm]{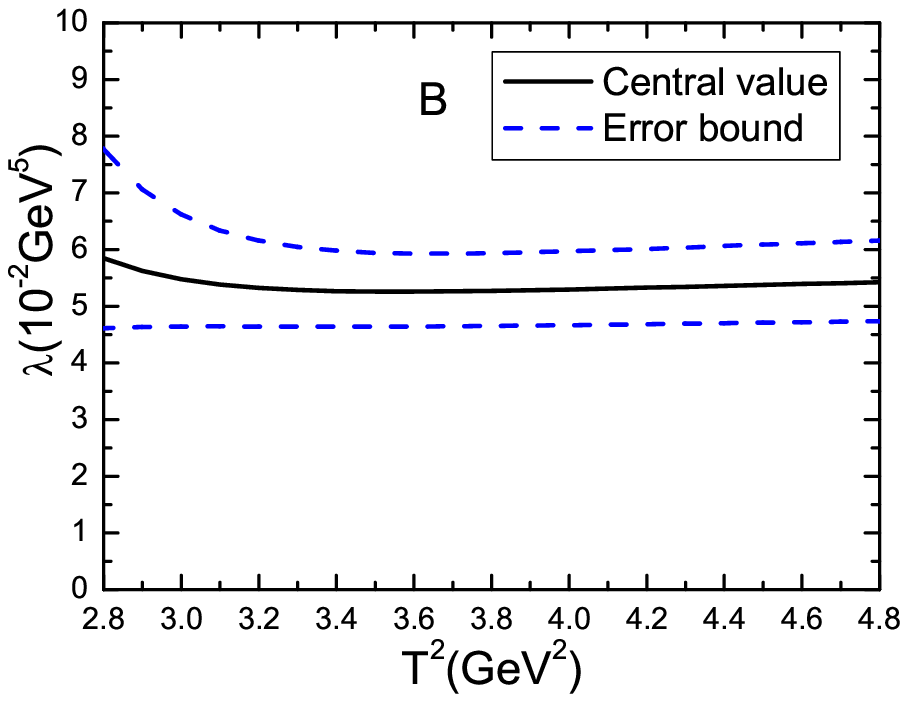}
\includegraphics[totalheight=6cm,width=7cm]{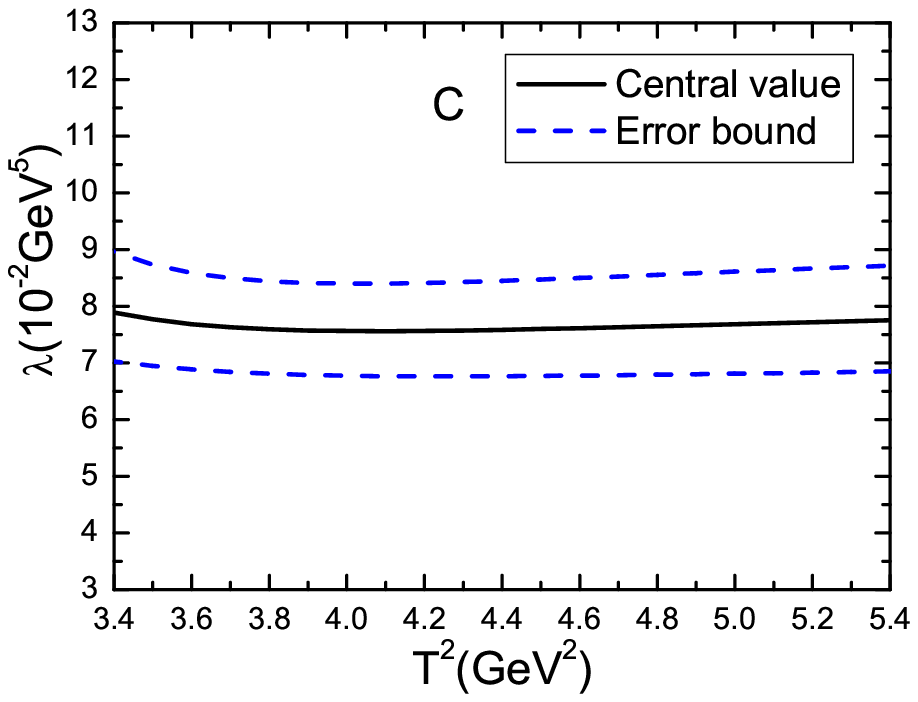}
\includegraphics[totalheight=6cm,width=7cm]{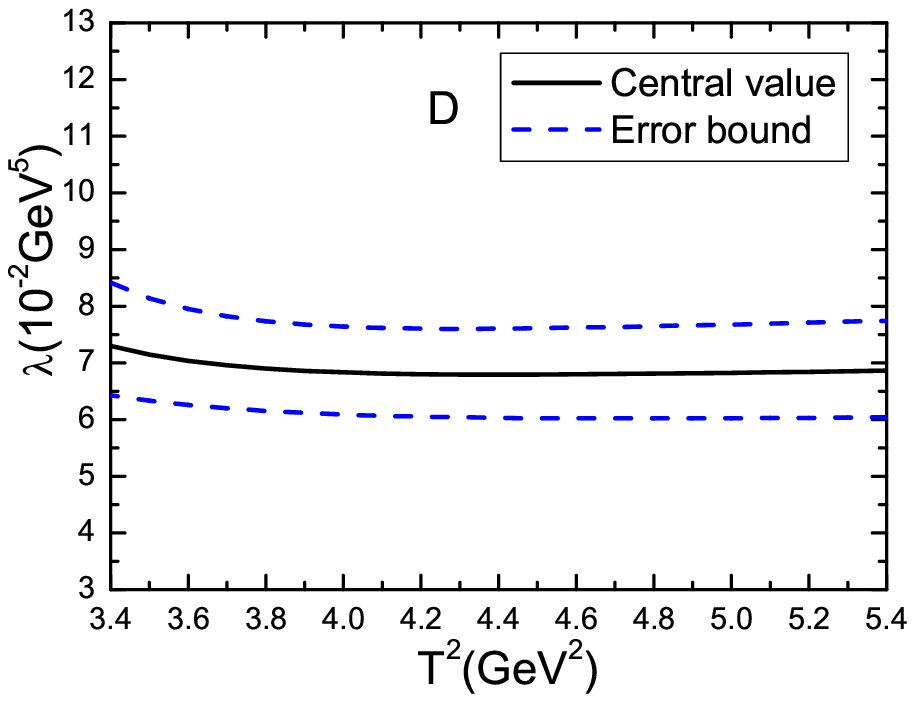}
  \caption{ The pole residues  with variations of the  Borel parameters $T^2$, where  the $A$, $B$, $C$ and $D$ denote the molecular states
  $D\bar{D}_1(1^{--})$, $D\bar{D}_1(1^{-+})$, $D^*\bar{D}_0^*(1^{--})$ and $D^*\bar{D}_0^*(1^{-+})$, respectively. }
\end{figure}

We take into account all uncertainties of the input parameters,
and obtain the values of the masses and pole residues of
 the   vector molecular states, which are  shown explicitly in Figs.1-2 and Table 1. In Figs.1-2,  we plot the masses and pole residues with variations
of the Borel parameters at much larger intervals   than the  Borel windows shown in Table 1. From the figures, we can see that there appear platforms  indeed in the Borel windows. Furthermore, from Table 1, we can see that the energy scale formula is well satisfied. Now the four criteria of the QCD sum rules are all satisfied, so we expect to make reasonable predictions.

The prediction  $M_{D\bar{D}_1(1^{--})}=4.36\pm0.08\,\rm{GeV}$ is consistent with the experimental data $M_{Y(4390)}=4391.6\pm6.3\pm1.0\,\rm{MeV}$ within uncertainties \cite{BES-Y4390}, while the predictions  $M_{D\bar{D}_1(1^{-+})}=4.60\pm 0.08\,\rm{GeV}$, $M_{D^*\bar{D}_0^*(1^{--})}=4.78\pm0.07\,\rm{GeV}$ and $M_{D^*\bar{D}_0^*(1^{-+})}=4.73\pm0.07\,\rm{GeV}$   are much larger than upper bound of the experimental data $M_{Y(4390)}=4218.4\pm4.0\pm0.9\,\rm{MeV}$ and $M_{Y(4390)}=4391.6\pm6.3\pm1.0\,\rm{MeV}$ \cite{BES-Y4390}, moreover, they are much larger than the near thresholds $M_{D^+D_1(2420)^{-}}=4293\,\rm{MeV}$, $M_{D^0D_1(2420)^{0}}=4285\,\rm{MeV}$, $M_{D^{*+}D_0^{*}(2400)^-}=4361\,\rm{MeV}$,   $M_{D^{*0}D_0^{*}(2400)^0}=4325\,\rm{MeV}$ \cite{PDG}.
The present predictions only favor assigning the $Y(4390)$   to be the $D\bar{D}_1(1^{--})$ molecular state.

In Ref.\cite{ZhangHuang-2009}, Zhang and Huang do not distinguish the charge conjugations and obtain the masses $M_{D^*\bar{D}^*_0}=4.26 \pm 0.07\,\rm{GeV}$ and $M_{D\bar{D}_1}=4.34 \pm 0.07\,\rm{GeV}$.
In Ref.\cite{Lee-Nielsen},  Lee, Morita and Nielsen distinguish the  charge conjugations and obtain the mass   $M_{D\bar{D}_1(1^{-+})}=4.19 \pm 0.22\,\rm{GeV}$.
In this article, we distinguish
the charge conjugations of the currents, calculate the contributions of the vacuum condensates up to dimension-10 in the operator product expansion in a consistent way, the intervals  of the vacuum condensates are much larger than the ones in Refs.\cite{ZhangHuang-2009,Lee-Nielsen}. Moreover, we   use the energy scale formula to determine the energy scales of the QCD spectral densities, which works well in our previous works \cite{WangHuang-molecule,Wang-molecule}. We obtain the predications  $M_{D\bar{D}_1(1^{--})}=4.36\pm0.08\,\rm{GeV}$, $M_{D\bar{D}_1(1^{-+})}=4.60\pm 0.08\,\rm{GeV}$, $M_{D^*\bar{D}_0^*(1^{--})}=4.78\pm0.07\,\rm{GeV}$ and $M_{D^*\bar{D}_0^*(1^{-+})}=4.73\pm0.07\,\rm{GeV}$, which differ from the ones in Refs.\cite{ZhangHuang-2009,Lee-Nielsen} significantly, the conclusion is changed.

In 2007, the  Belle collaboration  measured  the cross section for the process $e^+e^- \to \pi^+ \pi^- \psi^{\prime}$, and  observed two structures $Y(4360)$ and $Y(4660)$ in the $\pi^+ \pi^- \psi^{\prime}$  mass spectrum  at $(4361\pm 9\pm 9)\, \rm{MeV}$ with a width of \, $(74\pm 15\pm 10)\,\rm{ MeV}$ and   $(4664\pm 11\pm 5)\,\rm{ MeV}$ with a width of \, $(48\pm 15\pm 3) \,\rm{MeV}$, respectively \cite{Belle4660-0707}. The quantum numbers of the $Y(4360)$ and $Y(4660)$ are $J^{PC}=1^{--}$ \cite{PDG}. The $Y(4390)$ and $Y(4360)$ have analogous masses and widths, so they may be the same particle, the $D\bar{D}_1(1^{--})$ molecular state.
The main decay modes the $D\bar{D}_1(1^{--})$ molecular state  are $DD^*\pi$  \cite{ZhaoQ}, it is important to search for the decay modes $DD^*\pi$  to diagnose the nature of the $Y(4260)$ and $Y(4390)$.

\section{Conclusion}
In this article, we assign the $Y(4390)$ and  $Y(4220)$ to be the vector molecular states $D\bar{D}_1(2420)$ and $D^*\bar{D}_0^*(2400)$, respectively,   distinguish
the charge conjugations, construct   the color singlet-singlet type   currents   to interpolate  them.
We calculate the contributions of the vacuum condensates up to dimension-10 in the operator product expansion in a consistent way, use the energy scale formula to determine the energy scales of the QCD spectral densities, and study the masses and pole residues with the QCD sum rules in details. The present predictions only favor assigning the $Y(4390)$  to be the $D\bar{D}_1(1^{--})$ molecular state.

\section*{Appendix}
The explicit expressions of the QCD spectral densities $\rho(s,r)$,
\begin{eqnarray}
\rho(s,r)&=&\frac{1}{4096\pi^6}\int_{y_i}^{y_f}dy \int_{z_i}^{1-y}dz \, yz\,(1-y-z)^3\left(s-\overline{m}_c^2\right)^2\left(35s^2-26s\overline{m}_c^2+3\overline{m}_c^4 \right) \nonumber\\
&&+r\frac{m_c^2}{2048\pi^6}\int_{y_i}^{y_f}dy \int_{z_i}^{1-y}dz \,  (1-y-z)^3\left(s-\overline{m}_c^2\right)^3  \nonumber\\
&&+(1-r)\frac{3m_c\langle\bar{q}q\rangle}{256\pi^4}\int_{y_i}^{y_f}dy \int_{z_i}^{1-y}dz\,\left( y+z\right)\left(1-y-z \right)\left(s-\overline{m}_c^2\right)^2  \nonumber\\
&&-\frac{m_c^2}{3072\pi^4}\langle\frac{\alpha_sGG}{\pi}\rangle\int_{y_i}^{y_f}dy \int_{z_i}^{1-y}dz \, \left(\frac{y}{z^2}+\frac{z}{y^2} \right)\left(1-y-z\right)^3 \left\{ 8s-3\overline{m}_c^2+s^2\,\delta\left(s-\overline{m}_c^2\right)\right\} \nonumber\\
&&-r\frac{m_c^4}{6144\pi^4}\langle\frac{\alpha_sGG}{\pi}\rangle\int_{y_i}^{y_f}dy \int_{z_i}^{1-y}dz \, \left(\frac{1}{y^3}+\frac{1}{z^3} \right)\left(1-y-z\right)^3
\nonumber\\
&&+r\frac{m_c^2}{2048\pi^4}\langle\frac{\alpha_sGG}{\pi}\rangle\int_{y_i}^{y_f}dy \int_{z_i}^{1-y}dz \, \left(\frac{1}{y^2}+\frac{1}{z^2} \right)\left(1-y-z\right)^3 \left(s-\overline{m}_c^2\right) \nonumber\\
&&+\frac{1}{1024\pi^4}\langle\frac{\alpha_sGG}{\pi}\rangle\int_{y_i}^{y_f}dy \int_{z_i}^{1-y}dz \, \left(y+z\right)\left(1-y-z\right)^2\,s\,\left(5s-4\overline{m}_c^2\right) \nonumber\\
&&+r\frac{3m_c^2}{2048\pi^4}\langle\frac{\alpha_sGG}{\pi}\rangle\int_{y_i}^{y_f}dy \int_{z_i}^{1-y}dz \, \left(\frac{1}{y}+\frac{1}{z}\right)\left(1-y-z\right)^2\,\left(s-\overline{m}_c^2\right) \nonumber\\
&&-(1-r)\frac{3m_c\langle\bar{q}g_s\sigma Gq\rangle}{512\pi^4}\int_{y_i}^{y_f}dy \int_{z_i}^{1-y}dz\,\left( y+z\right)\left(s-\overline{m}_c^2\right)  \nonumber\\
&&-\frac{3m_c\langle\bar{q}g_s\sigma Gq\rangle}{256\pi^4}\int_{y_i}^{y_f}dy \int_{z_i}^{1-y}dz \,\left( \frac{y}{z}+\frac{z}{y}\right)\left( 1-y-z\right)\left(2s-\overline{m}_c^2\right)\nonumber\\
&&-r\frac{3m_c\langle\bar{q}g_s\sigma Gq\rangle}{256\pi^4}\int_{y_i}^{y_f}dy \int_{z_i}^{1-y}dz \, \left( 1-y-z\right)\left(s-\overline{m}_c^2\right)\nonumber\\
&&+r\frac{ \langle\bar{q}q\rangle^2}{32\pi^2}\int_{y_i}^{y_f}dy\,y(1-y)\,\left(s-\widetilde{m}_c^2-2r\widetilde{m}_c^2\right)\nonumber\\
&&+\frac{g_s^2\langle\bar{q}q\rangle^2}{864\pi^4}\int_{y_i}^{y_f}dy \int_{z_i}^{1-y}dz\,yz\,\left\{8s-3\overline{m}_c^2 +r\frac{m_c^2}{2yz}+s^2\,\delta\left(s-\overline{m}_c^2 \right)\right\}\nonumber\\
&&-\frac{g_s^2\langle\bar{q}q\rangle^2}{1728\pi^4}\int_{y_i}^{y_f}dy \int_{z_i}^{1-y}dz\,\left(1-y-z\right)\,\left\{3\left( \frac{y}{z}+\frac{z}{y}\right)\left(7s-4\overline{m}_c^2 \right) +\left(\frac{y}{z^2}+\frac{z}{y^2}\right)m_c^2\right.\nonumber\\
&&\left. \left[7+5s\,\delta\left(s-\overline{m}_c^2 \right)\right]-\left( y+z\right)\left(4s-3\overline{m}_c^2 \right)+r\left(\frac{1}{y}+\frac{1}{z}\right)\frac{3m_c^2}{2}\right\} \nonumber\\
&&-(1-r)\frac{m_c^3\langle\bar{q}q\rangle}{768\pi^2}\langle\frac{\alpha_sGG}{\pi}\rangle\int_{y_i}^{y_f}dy \int_{z_i}^{1-y}dz\,\left(y+z\right)\left( 1-y-z\right)\left( \frac{1}{y^3}+\frac{1}{z^3}\right)\,\delta\left(s-\overline{m}_c^2\right) \nonumber\\
&&+(1-r)\frac{m_c\langle\bar{q}q\rangle}{256\pi^2}\langle\frac{\alpha_sGG}{\pi}\rangle\int_{y_i}^{y_f}dy \int_{z_i}^{1-y}dz\, \left( 1-y-z\right)\left( \frac{z}{y^2}+\frac{y}{z^2}\right)  \nonumber\\
&&+\frac{3m_c\langle\bar{q}q\rangle}{128\pi^2}\langle\frac{\alpha_sGG}{\pi}\rangle\int_{y_i}^{y_f}dy \int_{z_i}^{1-y}dz \, \left\{1+\frac{4s}{9}\,\delta\left(s-\overline{m}_c^2\right) \right\} \nonumber\\
&&-r\frac{m_c\langle\bar{q}q\rangle}{256\pi^2}\langle\frac{\alpha_sGG}{\pi}\rangle\int_{y_i}^{y_f}dy \int_{z_i}^{1-y}dz \, \left\{\left(\frac{y}{z}+\frac{z}{y}\right)-\frac{r(1-r)}{6} \right\}\nonumber\\
&&+\frac{ m_c^2\langle\bar{q}q\rangle\langle\bar{q}g_s\sigma Gq\rangle }{32\pi^2}\int_{y_i}^{y_f}dy\,\left(1+\frac{s}{T^2}\right)\,\delta\left(s-\widetilde{m}_c^2\right)\nonumber
\end{eqnarray}

\begin{eqnarray}
&&-r\frac{ 3\langle\bar{q}q\rangle\langle\bar{q}g_s\sigma Gq\rangle }{64\pi^2}\int_{y_i}^{y_f}dy\,y(1-y)\,\left\{1+\frac{s}{3}\,\delta\left(s-\widetilde{m}_c^2\right)\right\}\nonumber\\
&&+r\frac{ \langle\bar{q}q\rangle\langle\bar{q}g_s\sigma Gq\rangle}{64\pi^2}\int_{y_i}^{y_f}dy \left\{\frac{1}{2}- r s\,\delta\left(s-\widetilde{m}_c^2\right) \right\}\nonumber\\
&&+r\frac{  \langle\bar{q}g_s\sigma Gq\rangle^2 }{256\pi^2}\int_{y_i}^{y_f}dy\,y(1-y)\,\left(3+\frac{2s}{T^2}+\frac{s^2}{2T^4}-r\frac{s^3}{T^6}\right)\,\delta\left(s-\widetilde{m}_c^2\right)\nonumber\\
&&+\frac{m_c^4\langle\bar{q}q\rangle^2}{288T^4}\langle\frac{\alpha_sGG}{\pi}\rangle\int_{y_i}^{y_f}dy   \, \left\{\frac{1}{y^3}+\frac{1}{(1-y)^3} \right\} \,\delta\left(s-\widetilde{m}_c^2\right) \nonumber\\
&&-r\frac{m_c^2\langle\bar{q}q\rangle^2}{576T^2}\langle\frac{\alpha_sGG}{\pi}\rangle\int_{y_i}^{y_f}dy   \, \left\{\frac{1-y}{y^2}+\frac{y}{(1-y)^2} \right\} \,\delta\left(s-\widetilde{m}_c^2\right) \nonumber\\
&&-\frac{m_c^2\langle\bar{q}q\rangle^2}{96T^2}\langle\frac{\alpha_sGG}{\pi}\rangle\int_{y_i}^{y_f}dy   \, \left\{\frac{1}{y^2}+\frac{1}{(1-y)^2} \right\} \,\delta\left(s-\widetilde{m}_c^2\right) \nonumber\\
&&-r\frac{  \langle\bar{q}g_s\sigma Gq\rangle^2}{512\pi^2}\int_{y_i}^{y_f}dy  \, \left( \frac{26}{27}+\frac{s}{T^2}-r\frac{2s^2}{T^4}\right) \,\delta\left(s-\widetilde{m}_c^2\right)\nonumber\\
&&+r\frac{\langle\bar{q}q\rangle^2}{96}\langle\frac{\alpha_sGG}{\pi}\rangle\int_{y_i}^{y_f}dy \,y(1-y) \,\left(1+\frac{2s}{3T^2}+\frac{s^2}{6T^4} -r\frac{s^3}{3T^6}\right)\,\delta\left(s-\widetilde{m}_c^2 \right)\, ,
\end{eqnarray}
where $y_{f}=\frac{1+\sqrt{1-4m_c^2/s}}{2}$,
$y_{i}=\frac{1-\sqrt{1-4m_c^2/s}}{2}$, $z_{i}=\frac{ym_c^2}{y s -m_c^2}$, $\overline{m}_c^2=\frac{(y+z)m_c^2}{yz}$,
$ \widetilde{m}_c^2=\frac{m_c^2}{y(1-y)}$, $\int_{y_i}^{y_f}dy \to \int_{0}^{1}dy$, $\int_{z_i}^{1-y}dz \to \int_{0}^{1-y}dz$ when the $\delta$ functions $\delta\left(s-\overline{m}_c^2\right)$ and $\delta\left(s-\widetilde{m}_c^2\right)$ appear.

\section*{Acknowledgements}
This  work is supported by National Natural Science Foundation, Grant Number 11375063.

\end{document}